\documentclass[aps,prb,twocolumn,showpacs]{revtex4}

\usepackage{graphicx}
\DeclareGraphicsExtensions{.png,.jpg,.eps}
\usepackage{multirow}
\usepackage{braket}
\usepackage{xcolor}
\usepackage{amsmath}

\begin{document}

\title{Optimizing tip-surface interactions in ESR-STM experiments}

\author{S. A. Rodr\'iguez$^{1}$, S. S.  G\'omez$^{1}$, J. Fern\'andez-Rossier$^{2}$
\footnote{On leave from Departamento de Fisica Aplicada, Universidad de Alicante, 03690 Spain },
A. Ferr\'on$^{1}$
}
\affiliation{ 
(1) Instituto de Modelado e Innovaci\'on Tecnol\'ogica (CONICET-UNNE) and 
Facultad de Ciencias Exactas, Naturales y Agrimensura, Universidad Nacional 
del Nordeste, Avenida Libertad 5400, W3404AAS Corrientes, Argentina.
\\(2) International Iberian Nanotechnology Laboratory (INL),
Av. Mestre Jos\'e Veiga, 4715-330 Braga, Portugal. 
}

\date{\today}

\begin{abstract}
Electron-spin resonance carried out with scanning tunneling microscopes (ESR-STM) is a recently developed experimental technique that is attracting enormous interest on account of its potential to carry out single-spin on-surface resonance with subatomic resolution. Here we carry out a theoretical study of the role of  tip-adatom interactions and provide guidelines for choosing the experimental parameters in order to optimize spin resonance measurements. We consider  the case of the Fe adatom on a MgO surface and its interaction with the spin-polarized STM  tip. We address three problems: first, how to optimize the tip-sample distance to  cancel the effective magnetic field created by the tip on the surface spin, in order to carry out proper magnetic field sensing.  Second, how to reduce the voltage dependence of the surface-spin resonant frequency, in order to minimize tip-induced decoherence due to voltage noise.  Third, we propose an experimental protocol to infer the detuning angle between the applied field and the tip magnetization, which plays a crucial   role in the modeling of the experimental results.  
\end{abstract}

\maketitle
\section{Introduction}\label{s1}

Properly designed experiments seek to  minimize the influence of the probing apparatus on the system of interest.
This becomes increasingly difficult  at the nanoscale, where macroscopic instruments interact with nanometric systems, and particularly challenging
when it comes to probe quantum systems.  In this work, we address this issue in the context of electron spin resonance (ESR)
driven with a scanning tunneling microscope (STM). After several decades of attempts \cite{manassen89,
reviewbalatsky2012},  reproducible ESR-STM 
of individual adatoms on a surface of MgO(100)/Ag  was reported in 2015\cite{Baumann2015}. This  has
paved the way for many other outstanding advances in the study of spin physics of individual magnetic atoms\cite{natterer2017,choi2017,yang2017,willke2018,
willke2018b,bae2018,yang2018,yang2019,willke2019,willke2019b}.
Spin resonance of isolated magnetic atoms promises novel applications ranging from quantum information technology to atomic-scale magnetometry.

ESR-STM has now been implemented in several different labs, extending the temperature range, both down to the mili-Kelvin regime\cite{Akaje2021,steinbrecher21}, as well 
as towards higher temperatures\cite{natterer2019,Bae2022}. Recently  experiments with higher driving frequencies were performed \cite {seifert2019}. ESR-STM has been demonstrated now in individual atoms (Fe, Cu), hydrogenated Ti\cite{yang2017,yang2018,steinbrecher21,kot22}, both alone and in artificially created structures such as dimers\cite{yang2017,bae2018,veldman21,kot22}, trimers, and tetramers\cite{yang2021}, and on  
alkali atoms on MgO \cite{kovarik2022}, as well as molecules \cite{willke2021a,zhang2021}.
The state-of-the-art spectral resolution of this ESR-STM, down to a few MHz, makes it  possible to resolve the hyperfine structure of Fe, Ti, and Cu atoms\cite{willke2018b,yang2018,farinacci22}. ESR-STM setup can also be used to manipulate surface spins coherently\cite{yang2019b}, with pulses, as well as to   drive nuclear spin states \cite{yang2018}.

Different mechanisms can account for 
the driving of the surface spin by the tip bias voltage
\cite{reviewbalatsky2012,arrachea2014,berggren2016,lado2017,
shakirov2019,galvez2019,ferron2019,delgado21}. For instance, in Ref.
[\onlinecite{lado2017}] two of us proposed a  mechanism  based on the modulation of
the exchange interaction between the magnetic tip and the magnetic adatom,
that originates from the piezoelectric distortion of the adatom.
In [\onlinecite{ferron2019}]  we proposed another complementary mechanism, that
can coexist with the others, based on the electric modulation of the $g$
tensor associated with the piezoelectric distortion of the adatom.

Spin interactions between the tip and on-surface species are definitely needed for the detection of ESR-STM, as the resonance readout is magnetoresistive, but they also bring unwanted features, such as uncontrolled variations of the local magnetic field of the surface spins that make absolute magnetometry measurements difficult and may also induce dephasing on the surface spin as mechanical and electrical noise lead to spin noise. Expectedly,  \cite{lado2017,willke2019,seifert2021} the magnetic interaction 
between the tip and the surface spins strongly depends on their separation, the nature of the ad-atom,  the tip design, and the angle formed by the tip magnetization vector and the applied field, that is non-zero on account of the tip magnetic anisotropy. Importantly, the tip-atom distance is expected to depend on the DC voltage drop at the STM-surface junction, on account of the piezoelectric displacement of the surface spins\cite{lado2017,seifert2020}. Therefore, we see the tip exerts an influence on the surface spins.

The main goal of the present work is to provide a theoretical basis that permits to control and quantify this influence, improving thereby the sensing capabilities of ESR-STM. Specifically, we consider the case of a single Fe atom on MgO and address three problems: First, we analyze  the optimal tip-adatom distance that minimizes the tip-induced effective magnetic field on the surface spin. This point is known as the No-tip influence (NOTIN) point $d_{n}$ and it was described by Seifert and co-workers in [\onlinecite{seifert2021}] as the  sweet spot where  the tip does not influence the  measured resonance frequency.    Second, we study  how to reduce the voltage dependence of the surface-spin resonance frequency and we show how this reduces   tip-induced decoherence due to voltage noise.  Third, we propose an experimental protocol to infer the detuning angle $\delta$ between the applied field and the tip magnetization, which plays a crucial   role in the modeling of the experimental results.

\begin{figure}[hbt]
\includegraphics[width=0.60\linewidth]{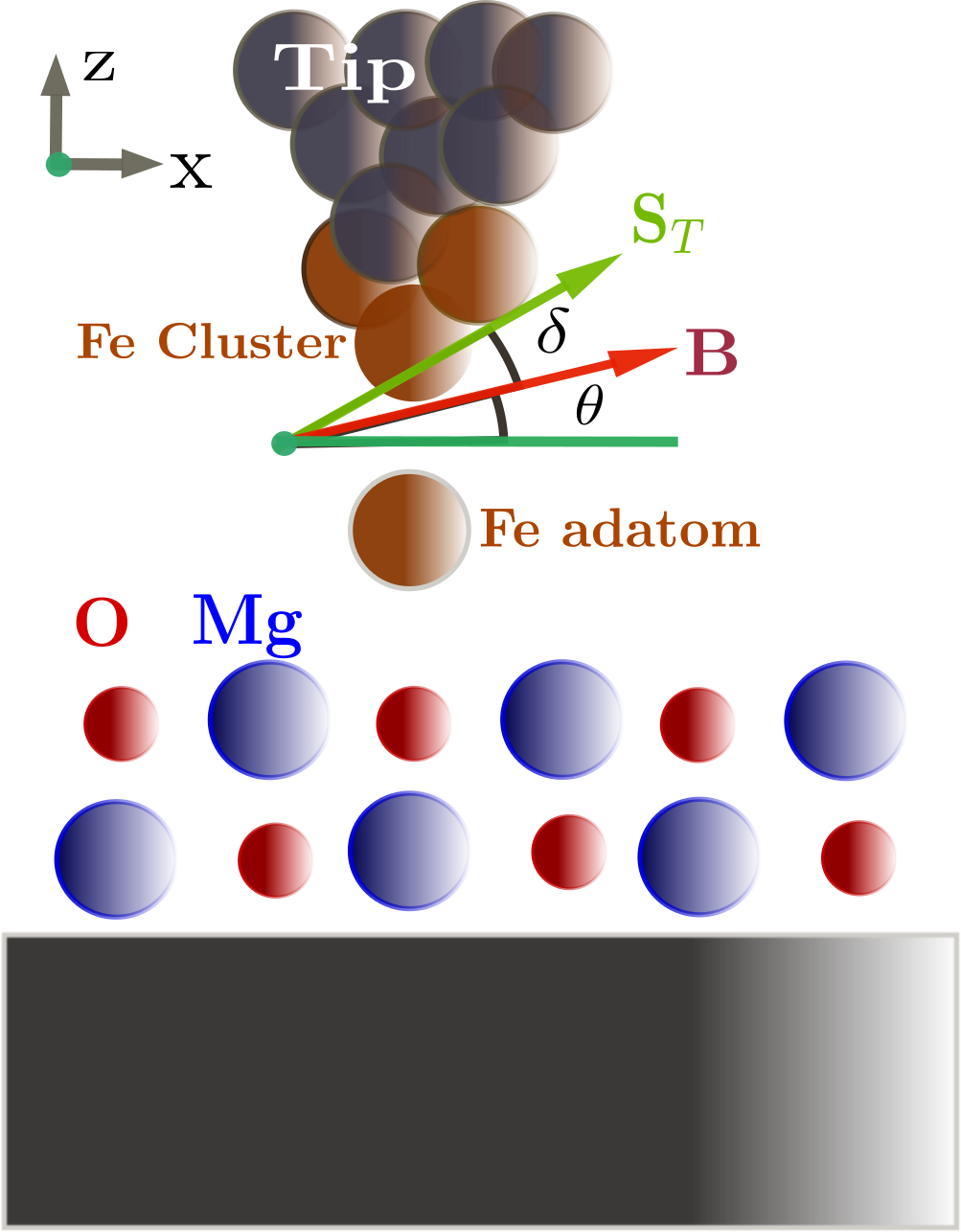}
\caption{\label{fig1} Sketch of a Fe atom in MgO/Ag and an SPSTM-tip }
\end{figure}

The rest of this paper is organized as follows. In sec. \ref{s2}, we present our models to address the problem of ESR-STM experiments in Fe at MgO. Sec. \ref{s3} is 
devoted to the study of the resonance frequency and the 
effect of the tip in ESR-STM experiments. We analyze, in detail, the possibility of optimizing the tip-sample interaction to perform ESR-STM experiments. In Sec. \ref{s4} we propose a simple method to determine the main magnetic properties of the STM-tip. Finally in Sec. \ref{sc} we  describe the most important findings and summarize the conclusions of the  work.


\section{Model Hamiltonian}\label{s2}

\subsection{Spin Hamiltonian}

The spin physics of  individual adatoms can be described at two different levels\cite{Abragam_Bleaney_book_1970}. First, a  multi-orbital electronic model for the outermost $d$  electrons, which includes Coulomb interactions, crystal field, and spin-orbit coupling, that we describe in Appendix A.  The ground state manifold, obtained by numerical diagonalization, can  be also described with an effective {\em spin model}.   For the case  of Fe on MgO the effective spin model, at zero field and ignoring coupling to the tip, is given by\cite{Baumann2015,baumann2015b,ferron2015,ferron2019}:
\begin{eqnarray}\label{hspin-g}
	\hat{H}_{eff}=-{\cal D}_2\hat{S}_z^2+{\cal D}_4\hat{S}_z^4-
	{\cal F}(\hat{S}_+^4+\hat{S}_-^4)
	\end{eqnarray}
 
\noindent where the spin operators act on the $S=2$ subspace. The anisotropy terms ${\cal D}_2$, ${\cal D}_4$ and ${\cal F}$ can be obtained from the diagonalization of the multi-orbital electronic model  as shown in the Appendix \ref{a2}. Now, the spectrum of the ground state manifold has an EPR active space formed by a doublet of states with $S_z=\pm 2$. Yet, this doublet has 
a zero-field splitting (ZFS), given by
$\Delta_{ZFS}= 48{\cal F}=0.2$ $\mu$eV, due to quantum spin tunneling
\cite{Klein_ajp_1951,garg1993topologically,Delgado_Loth_epl_2015}.
We treat tip spin as a classical unit vector, $\vec{n}_T$. As a result, the interaction of the adatom with the tip can be treated as an effective field:

\begin{equation}\label{h1j}
{\cal H}_1=	\mu_B\sum_{\beta=x,y,z} 
	g^{\beta}  B_{eff}^{\beta} S^{\beta}.
\end{equation}

\noindent If the magnetic field is along the $z$ direction, we find that the gap between the ground state and the first excited state is

\begin{equation}
	\Delta\approx2\sqrt{(24{\cal F})^2+(2\mu_B g^{z}  B_{eff}^{z})^2}\,.
	\label{gapp1}
\end{equation}

\noindent This equation is still true  if we add the   component of the magnetic field along the $x$ direction, as long as the Zeeman energy associated to $B_{eff}^{x}$ is smaller than {\bf ${\cal D}_2+{\cal D}_4$}, as we show in Appendix \ref{a2}. In the present case, the effective field $\vec{B}_{eff}$  is the sum of three contributions, external magnetic field, dipolar field of the tip, and exchange field of the tip:

\begin{equation}
B^{\beta}_{eff}=B^{\beta}+ \frac{J_{ex}(z)}{g^{\beta}\mu_B}n_T^{\beta}+B^{\beta}_{dip}
\equiv B^{\beta}+B^{\beta}_{\rm tip}
\label{bej}
\end{equation}

\noindent where  $B^{\beta}$ is the $\beta=x,y,z$ component of the external magnetic field, 

\begin{equation}
    J_{ex}(z)=\langle S_{tip}\rangle J_{0}\,e^{-z/l_{0}}
    \label{eq:jex}
\end{equation} 

\noindent is the distance-dependent tip-adatom exchange, $S_{tip}$ is the spin of the tip and $n_T^{\beta}$ are the components of the unit vector that describes the orientation of the tip spin. In the rest of the work, and so that our results show agreement with the experiments, we will use $J_0=20$ eV, $l_0=0.04$ nm and $\langle S_{tip}\rangle=2$.

\noindent Given the $C_4$ symmetry of the Fe adatom on top of an oxygen atom on the 001 MgO surface, we can assume the external field lies in the xz plane:
\begin{equation}
\vec{B}=B(\cos\theta,0,\sin\theta)
\end{equation}
Because of its magnetic anisotropy, the tip magnetization can be misaligned from the external field by an angle $\delta$:
 \begin{equation}
\vec{n}_T=(\cos(\theta+\delta),0,\sin(\theta+\delta))
\label{def:delta}
\end{equation}

The dipolar interaction 
between the magnetic moment of the tip and the 
surface spin comes from the magnetic field created by the tip,

\begin{equation}
    \vec{B}_{dip}=\frac{\mu_0}{4\pi}\left(3\frac{(\vec{m}\cdot\vec{d})\vec{d}}{d^5}-\frac{\vec{m}}{d^3}\right)
    \label{eq:dip}
\end{equation}

\noindent where $\vec{m}=g_{\rm tip}\mu_B \langle S_{tip}\rangle  \vec{n}_T\equiv M_{\rm tip} \vec{n}_T $.  Of course, $M_{\rm tip}$ depends on the number of atoms in the tip.  In the literature\cite{willke2019} values of $M_{\rm tip}\simeq 10\mu_B-40 \mu_B$ seems to be reasonable. In the rest of the work, and so that our results show agreement with some of the last experiments, we will use $M_{\rm tip}=30 \mu_B$.
When the Fe atom is just above the O atom we have $\vec{B}_{dip}=\,\mu_0\, M_{tip}
	\hat{n}_{dip}/(4\pi|z|^3)$, where 
	$\hat{n}_{dip}=(\cos(\theta+\delta),0,-2\,\sin(\theta+\delta))$.

Altogether, we arrive to the following expression for the  $z$ component of the effective field 
acting on the Fe adatom:
\begin{equation}
 B_{eff}^{(z)}= \left(\frac{J_{ex}(z)}{g^z\mu_B}   
    -2
\frac{    \mu_0\, M_{tip} }{4\pi|z|^3}
    \right)   \sin(\theta+\delta) + B\sin\theta
    \label{eq:Beffz}
\end{equation}

\subsection{Effect of the tip-electric field on the surface spin}
In actual EPR-STM experiments, there is a $DC$ bias, with amplitude $V_{DC}$, superimposed to the $AC$ bias $V_{RF}\sin{(\omega t)}$. Applying an
electric field induces a strain $\delta z$ of the bond between the Fe adatom and the oxygen atom underneath \cite{lado2017}. This leads to a modulation of the crystal field \cite{lado2017,ferron2019} and the magnetic field induced by the tip\cite{lado2017} (see Appendix {\ref{a4}}). The
electric field across the gap between the STM and the MgO surface, $E=V_{DC}/d$, where $d$ is the tip-MgO distance, induces a force on the adatom, $F=q V_{DC}/d$ on account of its charge $q$. This force is compensated by a restoring elastic force $F=-k\delta z$. Then, we can evaluate how much the Fe atom is displaced from its equilibrium position \cite{lado2017}:
\begin{equation}
        \delta z=\frac{q V_{DC}}{kd}
        \label{piezo}
\end{equation}

\noindent With this sign convention, increasing $V_{DC}$ leads to a stretching of the Fe-O bond and a reduction of the tip-Fe distance.
In the present case we have $q=2$ and, from DFT
calculations for Fe on MgO \cite{lado2017,ferron2019}, we obtain $k=600$ eV nm$^{-2}$. 
We find a modulation of the crystal field parameters in
 the multi-orbital fermionic model (see Appendix \ref{a1} and Appendix \ref{a4}) whose low energy states are described by   Eq. (\ref{hspin-g}). Using DFT calculations, we have computed this dependence
for Fe on MgO \cite{lado2017,ferron2019}. Numerical diagonalization of the multi-orbital model that takes into account the modulation of the crystal field parameters as we described in Appendix \ref{a4} gives:
\begin{eqnarray}
\label{spin_parameters_mod-1}
 && {\cal D}_i={\cal D}_{i_{eq}}+\alpha_{{\cal D}_i}\,\delta z \nonumber\\
 && {\cal F}={\cal F}_{eq}+\alpha_{{\cal F}}\,\delta z \nonumber\\
 && g^{\beta}=g^{\beta}_{eq}+\alpha_{g^{\beta}}\,\delta z 
\end{eqnarray}
\noindent where the equilibrium parameters are those corresponding to a zero electric field when the piezoelectric displacement is not present. The parameters $\alpha_X$ are defined and evaluated in Appendix \ref{a4} and account for the effect of the piezoelectric displacement on the atom-surface interaction.

The combination of the piezoelectric  distortion of the atom-tip distance due to the external electric field (Eq. (\ref{piezo})) and the  crystal field parameter modulation (Eq. (\ref{spin_parameters_mod-1})) lead also to a 
modulation of  the effective tip-induced field:
\begin{equation}
\frac{\partial B^{\beta}_{eff}}{\partial z}= \frac{\partial J_{ex}(z)}{\partial z}\frac{1}{g^{\beta}\mu_B}n^{\beta}
+\frac{\partial B^{\beta}_{dip}}{\partial z}-\frac{J_{exc}(z)}{(g^{\beta})^2\mu_B}\, \frac{\partial g^{\beta}}{\partial z}
\end{equation}
 Finally, we can express the shift in the adatom resonance frequency, $f=\Delta/h$, as follows:
\begin{equation}\label{deltaf}
   \delta f= \frac{\partial f}{\partial B^{\beta}_{eff}} \frac{\partial B^{\beta}_{eff}}{\partial z}\,\delta z+
   \frac{\partial f}{\partial {\cal F} } \frac{\partial {\cal F} }{\partial z}\,\delta z+\frac{\partial f}{\partial g^{\beta} } \frac{\partial  g^{\beta} }{\partial z}\,\delta z
\end{equation}

\noindent with $\beta=z$.

\section{Sweet spots for optimal tip-sample interactions}\label{s3}
We now study how to optimize experimental parameters in the ESR-STM setup, such as the tip-adatom distance $d_{\rm tip}$ and the DC voltage $V_{DC}$. Specifically, we devote ourselves to describing two special points of operation.  First,  the  optimal $d_{\rm tip}$ for which the effective magnetic field vanishes.  The second, the regions in the $d_{\rm tip},V_{DC}$ plane where the variation of the frequency with respect to  $V_{DC}$ is either large, affording voltage-controlled resonance frequency, or vanishing,  which will mitigate dephasing.

\subsection{No-tip influence distance}\label{s3a}

To establish to what extent the magnetic field induced by the tip affects the measurements, we must know how the resonance frequency, given in Eq. (\ref{gapp1}) behaves as a function
of tip-sample distance when we experiment with different STM tips.

\begin{figure}[hbt]
\includegraphics[width=0.8\linewidth]{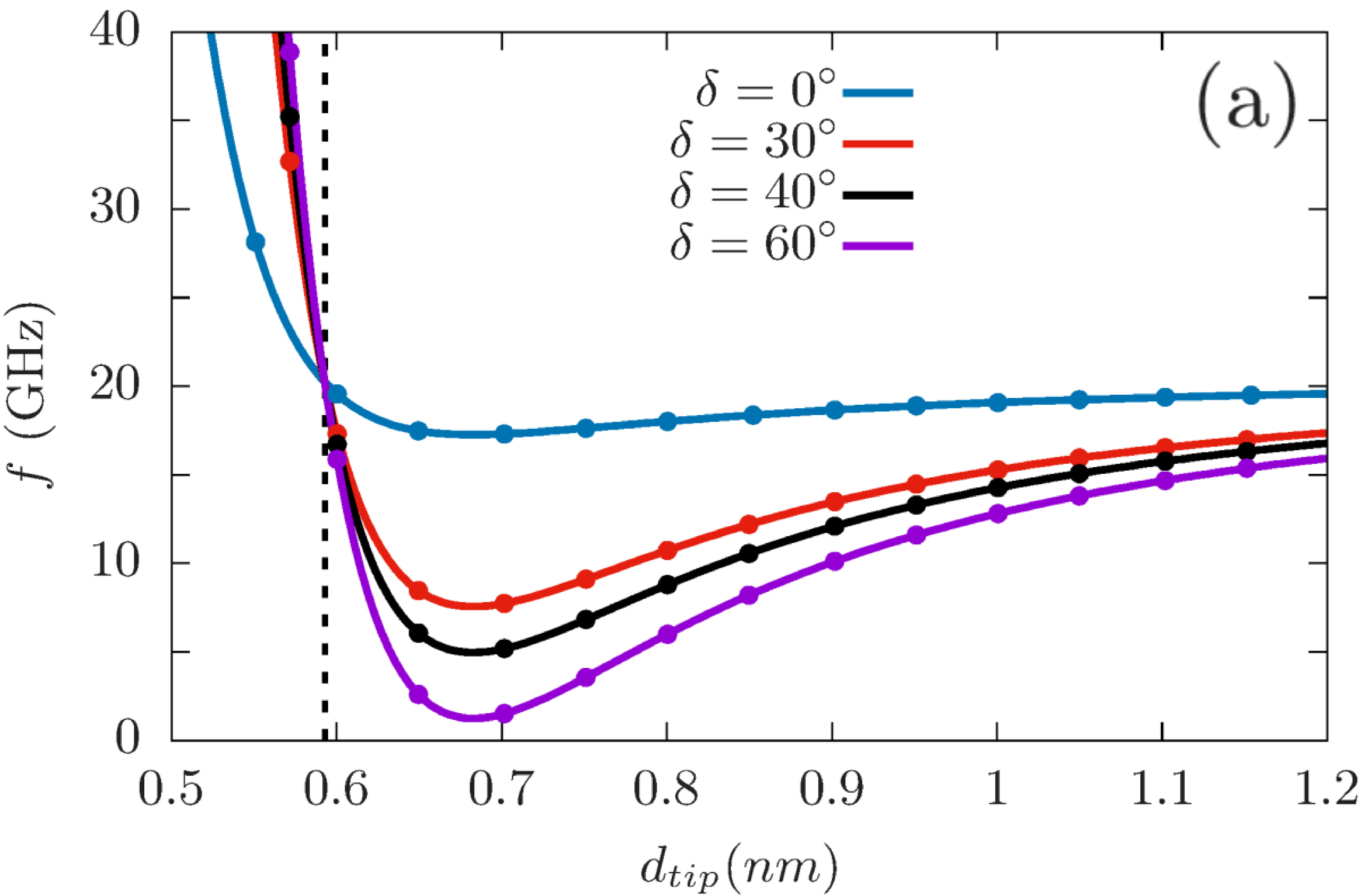}
\vspace{0.2cm}
\includegraphics[width=0.8\linewidth]{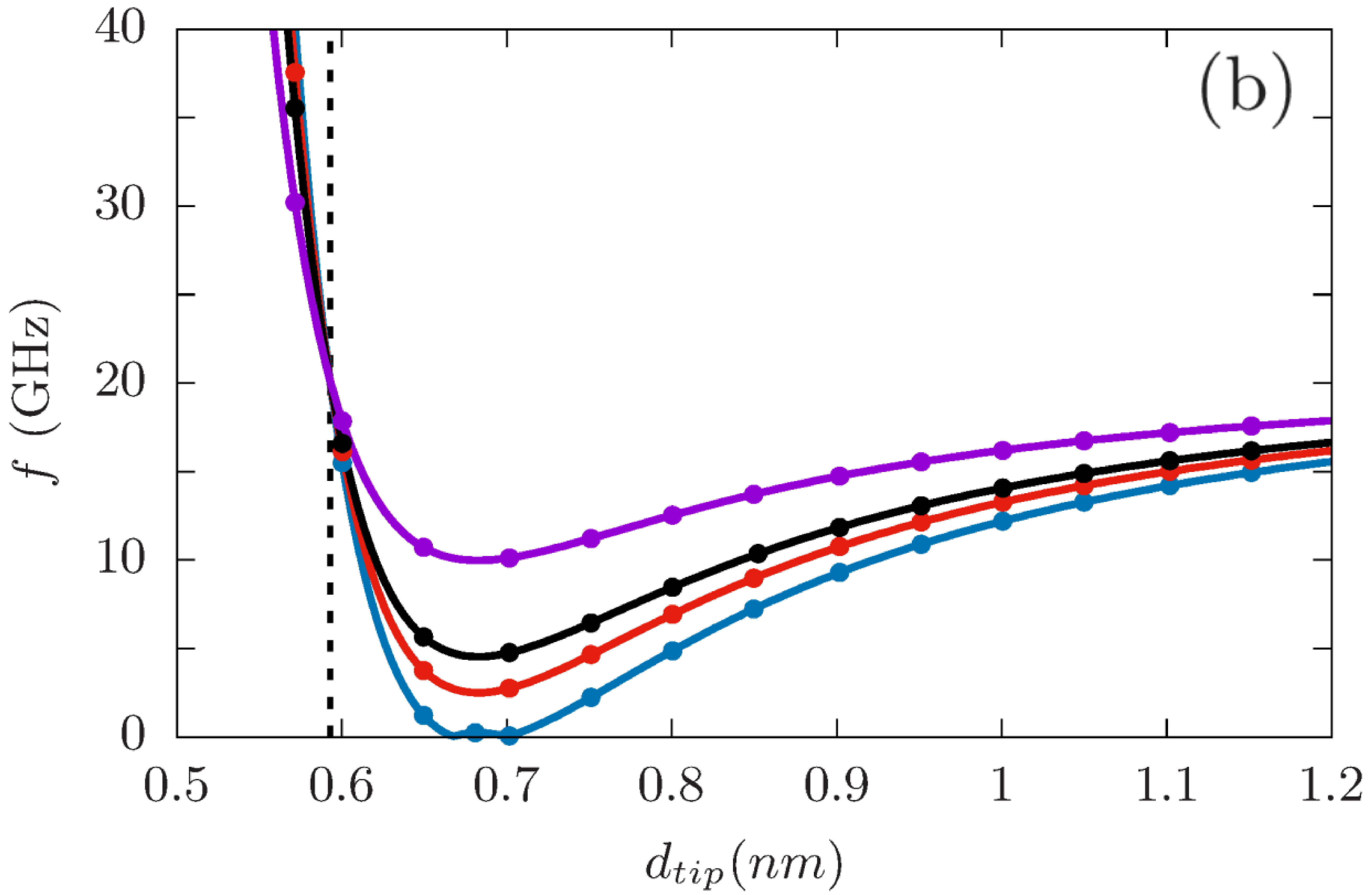}
\caption{\label{fig2}
Resonance frequency as a function of the
tip atom distance for different values of the tip anisotropy $\delta$. Panel (a) shows results for $\theta=8^{\circ}$ and $B_{ext}= 0.9$ T. Panel (b) corresponds to the results with $\theta=90^{\circ}$ and 
$B_{ext}= 0.1$ T. Solid lines show results obtained by direct diagonalization of the multi-orbital model while dots represent
perturbative results obtained using Eq. (\ref{gapp1}).}
\end{figure}

\noindent The influence of the tip arises from the $z$ component of the total magnetic field term (see Eq. (\ref{eq:Beffz})). It is important to note that ${\cal F}$ is small and therefore, in most situations, we can usually assume that $\Delta\simeq 4 \mu_B g_z B_{eff}{(z)}$. In Fig. \ref{fig2} we depict the ESR frequency as a function of the tip-adatom distance for different values of the tip anisotropy, $\delta$ (see Fig. \ref{fig1}). In solid lines,
we show the corresponding calculations obtained by direct diagonalization of our multi-orbital electronic model (see Appendix \ref{a1}) for two different experimental situations. Filled dots in Fig. \ref{fig2} show calculations performed using the perturbative expression Eq. (\ref{gapp1}). Panel (a) corresponds  to an external magnetic field almost in-plane \cite{willke2019,willke2018,willke2018b, willke2019b,yang2018,yang2017,yang2019} while in panel (b) an out-of-plane external magnetic field is applied \cite{seifert2020}. Expectedly, the resonance frequency goes to a plateau-like regime for a large atom-tip distance, as both exchange and dipolar interactions fade away. 

Interestingly,  we observe in Fig. \ref{fig2} a point at which all the curves intersect at a value approximately given by $4\mu_B g_z B_z$, independent of the dipolar and exchange field of the tip.
This happens when the off-plane component of the tip field vanishes,  ${B}_{tip}^{(z)}=0$ (see Eq. (\ref{bej}) and Eq. (\ref{eq:Beffz})). 
Using the expressions for the dipolar and exchange fields we obtain an implicit equation for the  tip-atom distance $d_n$ for which these two contributions cancel each other:

\begin{equation}\label{notin1JFR}
 J_0 e^{-d_{n}/l_0}= 
 \frac{g^z\gamma_tM_{tip}}{d_{n}^3},
 \end{equation}

\noindent where $\gamma_t=\mu_B\mu_0/ 4\pi$. For the choice of $J_0$, $l_0$ and $M_{tip}$, we find the NOTIN point at 
 $d_{n}=0.59$ nm. 

We observe, for the parameters employed
in Fig. \ref{fig2}, that displacements of $10$ pm from the NOTIN point
produce tip magnetic fields of $50$ mT. Although the NOTIN position seems to be optimal to carry out most of the measurements, it is clear that any small tip drift or tip vibration can make the scenario a little more complex by inducing unwanted magnetic fields and unexpected frequency alterations. For instance, the NOTIN distance has a small dependence on $V_{DC}$, by virtue of the piezoelectric displacement of the surface atom and the modulation of its $g$ factor, given in eq. (\ref{spin_parameters_mod-1}). For the chosen value of tip parameters and up  $V_{DC} = 200 mV$, we obtain changes around 2 pm in the NOTIN tip position.

\subsection{ Control of the $V_{DC}$ dependence }\label{s3b}

We now analyze how  DC electric  field of the tip  changes the resonance 
frequency of the surface spin. This dependence  arises from the combination of the piezoelectric displacement of the surface atom (Eq. \ref{piezo}) and the  distance dependence of the frequency (Eq. \ref{deltaf}). Depending on the target application, this dependence could be either a resource, for instance, to control $f$ without modifying the applied magnetic field, or a problem to avoid, as it can bring additional dephasing, as discussed below.   In order to quantify these matters, we define  the variation of the resonant frequency $f$ with respect to its $V_{DC}=0$ value,  
\begin{equation}
\delta f=f(V_{DC})-f(V_{DC}=0)
\label{deltaf2}
\end{equation} 

\begin{figure}[hbt]
\includegraphics[width=1.05\linewidth]{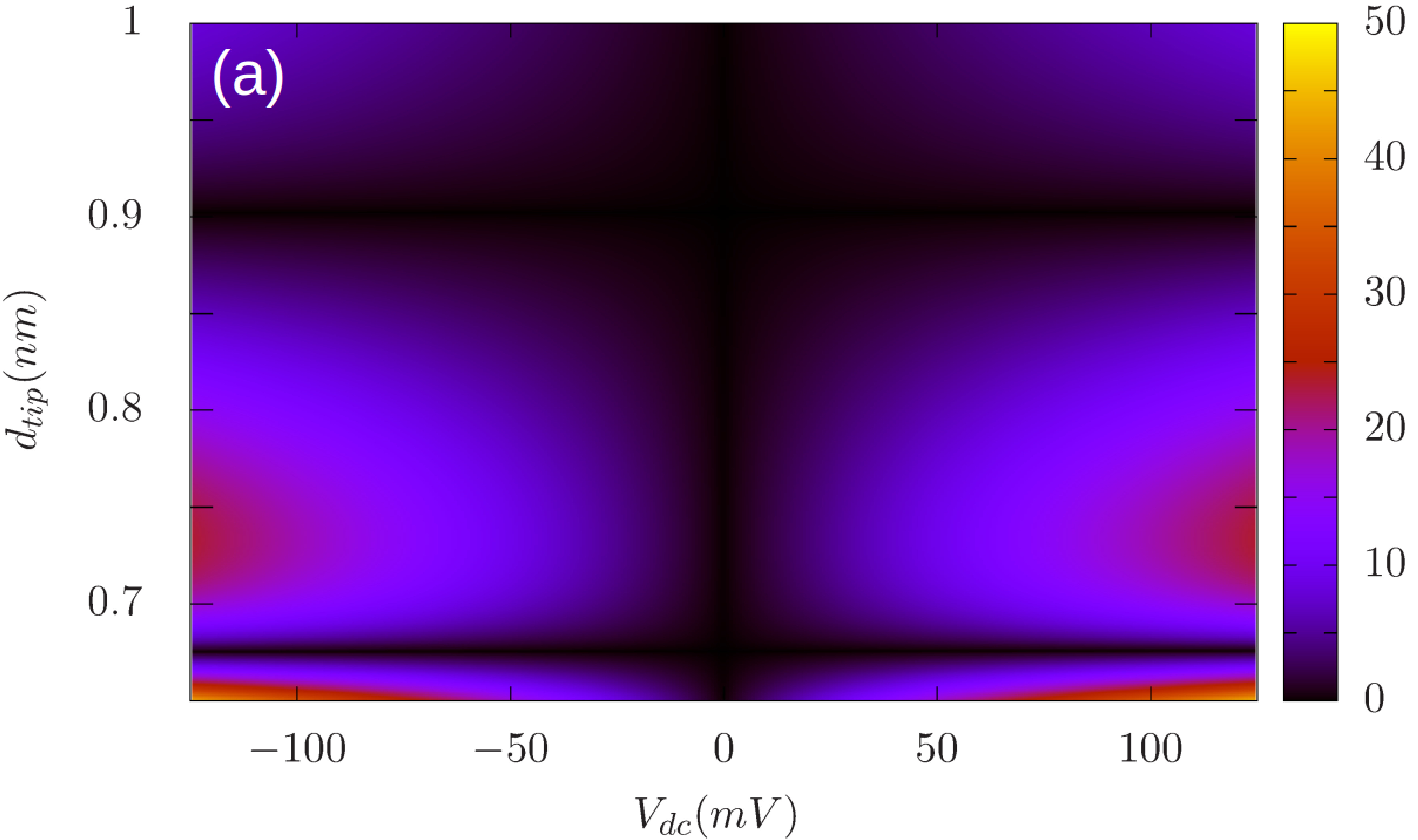}
\includegraphics[width=1.0\linewidth]{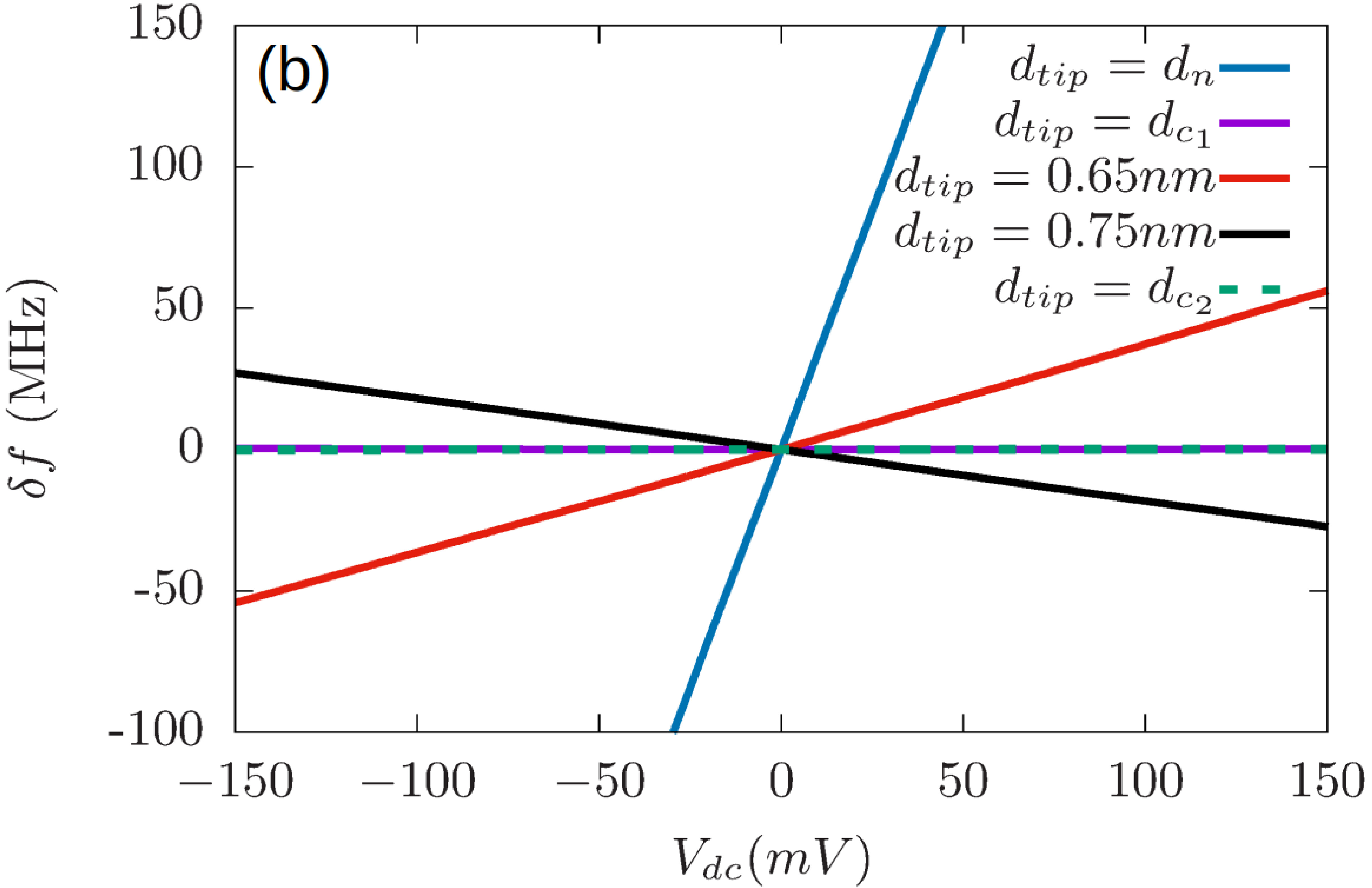}
\caption{\label{figsh}
Shift of the resonance frequency versus $V_{DC}$ (Eq. (\ref{deltaf2}) ). 
(a) Contour map of $|\delta f|$ as a function of the tip atom distance and external voltage for $\delta=60^\circ$ and $B=0.9$ T almost in-plane ($\theta=8^\circ$).
In Panel (b) we plot the shift $\delta f$ (a horizontal section of the map) as a function of the
external voltage for five different tip-atom distances.}
\end{figure}

The contour map of Fig.  \ref{figsh} (a) shows  $|\delta f|$ (Eq. \ref{deltaf2}) 
as a function of both  the external voltage and the tip-atom distance,
for $\theta=8^{\circ}$, $\delta=60^{\circ}$ and $B_{ext}= 0.9$ T.  We focus our attention first on regions where $\delta f$ is small.  This happens  both for $d\simeq 0.7$nm and $d\simeq 0.9$ nm.  In these two regions  we have  $df/dV_{DC}=0$ (see Eq. \ref{gapp1}) at $V_{DC}=0$.  

We now discuss how a small value of $df/dV_{DC}$  mitigates decoherence. It is known that, for  a two-level system, stochastic fluctuations of  fluctuating energy difference  lead to pure dephasing. Specifically, let us write the effective field  as the sum of the static contribution and a time-dependent fluctuation part, $ B_{eff}^z=B_{eff,0}^z+  b(t)$. For the  fluctuation function, we assume that the time average vanishes, $\overline{b(t)}=0$ but has finite short memory fluctuations over time, 
${\cal S}_b(t)\equiv \overline{b(t)b(t+\tau)}$ that decay rapidly 
 when $t>>\tau$, and have an amplitude that scale with $b_0^2$. We define the Fourier transform 
 $k(\omega)\equiv\frac{1}{2}\int_{-\infty}^\infty \overline{b(t)b(t+\tau)} e^{-i\omega t} dt$. 
We thus see that the stochastic field is characterized by the amplitude $b_0$ and a correlation time $\tau$. We can write down the decoherence rate as\cite{delgado2007review}:
\begin{equation}
    \frac{1}{T_2}= \frac{(2 g^z \mu_B)^2}{\hbar}  k(0) \propto \frac{(2 g^z \mu_B)^2}{\hbar}  b_0^2 \tau_0
\end{equation}
We note this dephasing mechanism is independent of the current-induced dephasing that has been observed experimentally\cite{willke2018}, and would act even if no current is tunneling through the surface spin. 
Now, it is apparent both electrical noise in $V_{DC}$, $\delta V_{DC}(t)$, and mechanical noise in $\delta z(t)$ will contribute to the amplitude of the fluctuations of the effective field: 
\begin{equation}
    b(t)= \frac{\partial B_{eff}^z}{\partial z} \left(\frac{\partial z}{\partial V_{DC}}
    \delta V_{DC}(t) + \delta z(t) \right)
\end{equation}
and thereby to dephasing.  Therefore, having both a small $df/dV_{DC}$ and a small $df/dz$ will eliminate this source of decoherence. Importantly, since the ESR-STM driving  of Fe on MgO is associated to the in-plane component of the effective field\cite{lado2017} and the frequency is dominated by the off-plane component,  it is possible to have a vanishing $df/dz$, and at the same time a large Rabi coupling.

We now focus on the regions where $df/dV_{DC}$ is not small so that the DC bias could be used to achieve
electrical control of the resonance frequency. In Fig. \ref{figsh} (b) we plot the change as a function of applied external voltage for five particular tip-atom distances.  For $d_{c1}\simeq 0.67$ nm (violet line) and $d_{c2}\simeq 0.9$ (cyan dashed line) we retrieve the stable regions discussed above, for which $f$ is independent of $V_{DC}$.  In contrast, for the other three values of $d_{tip}$ we find a strong dependence of $f$ on $V_{DC}$.  We draw attention to the immense value of $df/dV_{DC}$ at the NOTIN point.

\section{Determination of tip anisotropy}\label{s4}

So far, we have assumed that $\delta$, the misalignment angle  between the tip magnetic moment and the external magnetic field, is unknown and can take values in a rather wide  range, going from almost 0 $^\circ$ up to 60$^\circ$ \cite{yang2019,seifert2020}.
The origin of this misalignment is necessarily related to some type of tip  magnetic anisotropy 
that can arise  both from  anisotropy in the  $g$ factor  and   single-ion anisotropies. 

\begin{figure}[hbt]
\includegraphics[width=1.0\linewidth]{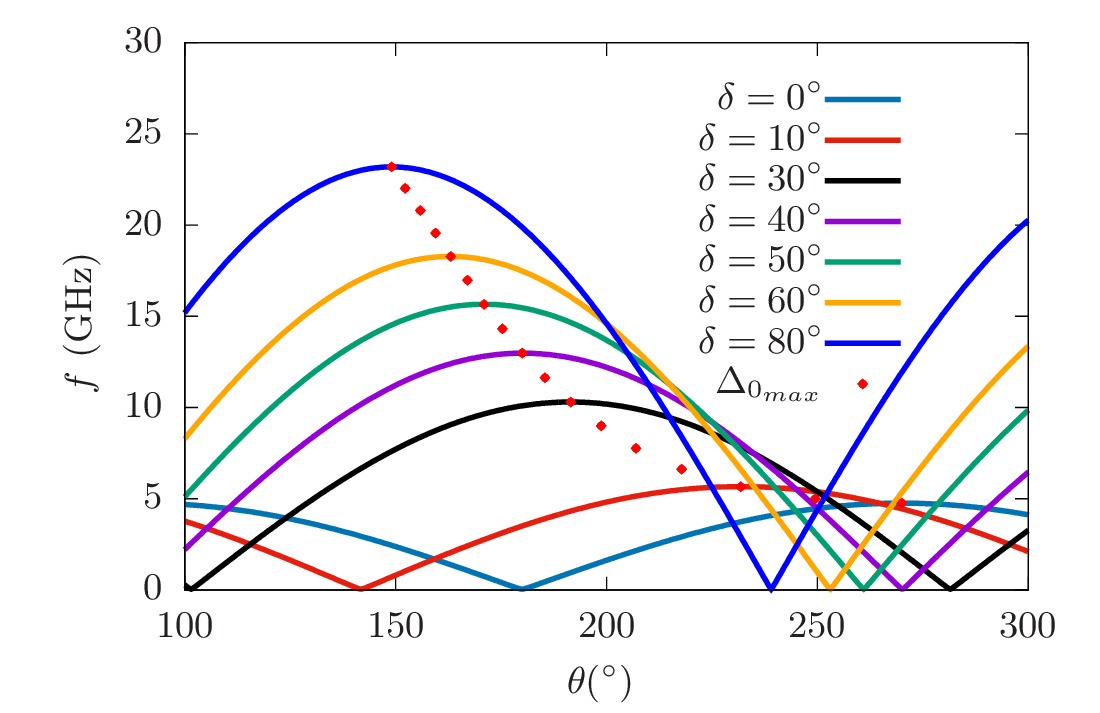}
\caption{\label{figtm1}
Resonance frequency as a function of the
external magnetic orientation, $\theta$, for different values of $\delta$. The calculation was performed by exact diagonalization of the few electron model (see appendix \ref{a1}). 
All the 
calculations were performed for $B=0.1$ T, $d_{tip}=0.7$ nm and a strong dipolar contribution $M_{tip}=30\,\mu_B$. 
}
\end{figure}

We now propose a simple experiment to determine $\delta$, similar to the one carried out by Kim {\em et al.}
\cite{kim2021} to probe the magnetic anisotropy of  hydrogenated Ti on the surface of MgO. Essentially, we propose to measure the resonant frequency for Fe on MgO  as a function of the angle $\theta$ formed between the external magnetic field and the surface, away from the NOTIN point.   Our calculations (see Fig. \ref{figtm1})
show a marked dependence of $f$ on $\theta$, expected because the Fe atom is sensitive mostly to the $z$ component of the effective field. We note how, for different values of $\delta$, our calculations show a lateral shift of the curves. Specifically, we find a dependence of the maximum as a function of the tip anisotropy $\delta$.  

Imposing that the  derivative of  Eq. (\ref{gapp1})  with respect to $\theta$ vanishes and using Eq. (\ref{eq:Beffz}), 
we obtain the following implicit equation for $\delta$


\begin{equation}\label{thmax}
 \tan(\theta_{max})=\frac{\varepsilon_{ext}+\varepsilon_{tip}\,\cos{(\delta)}}{\varepsilon_{tip}
	\,\sin{(\delta)}}\,.
\end{equation}

\noindent where

\begin{eqnarray}
\varepsilon_{ext}&=&\,g^z\mu_B\,B\, , \nonumber \\
\varepsilon_{tip}&=&\varepsilon_{exc}-\varepsilon_{dip} \, ,\nonumber \\
 \varepsilon_{exc}&=&2 J_{0}\,e^{-z/l_0} \, ,\nonumber \\
 \varepsilon_{dip}&=&2\gamma_t\frac{g^z\,M_{tip}}{|z|^3}\, ,
\end{eqnarray}

\noindent $g_z$ is defined in Eq. (\ref{spin_parameters_mod-1}) and $z=d_{\rm tip} -\delta z$.
Thus, experimental determination of $\theta_{max}$ and the tip effective fields, $\varepsilon_{tip}$ and $\varepsilon_{exc}$,
permit to read out the value of $\delta$.
Importantly, 
in the limit of a very small magnetic field, we have a simple linear relation between
$\delta$ and $\theta_{max}$, independent of $\varepsilon_{tip}$ and $\varepsilon_{exc}$,
\begin{equation}\label{tad0}
\delta=\frac{\pi}{2}-\theta_{max}. 
\end{equation}

\begin{figure}[hbt]
\includegraphics[width=0.950\linewidth]{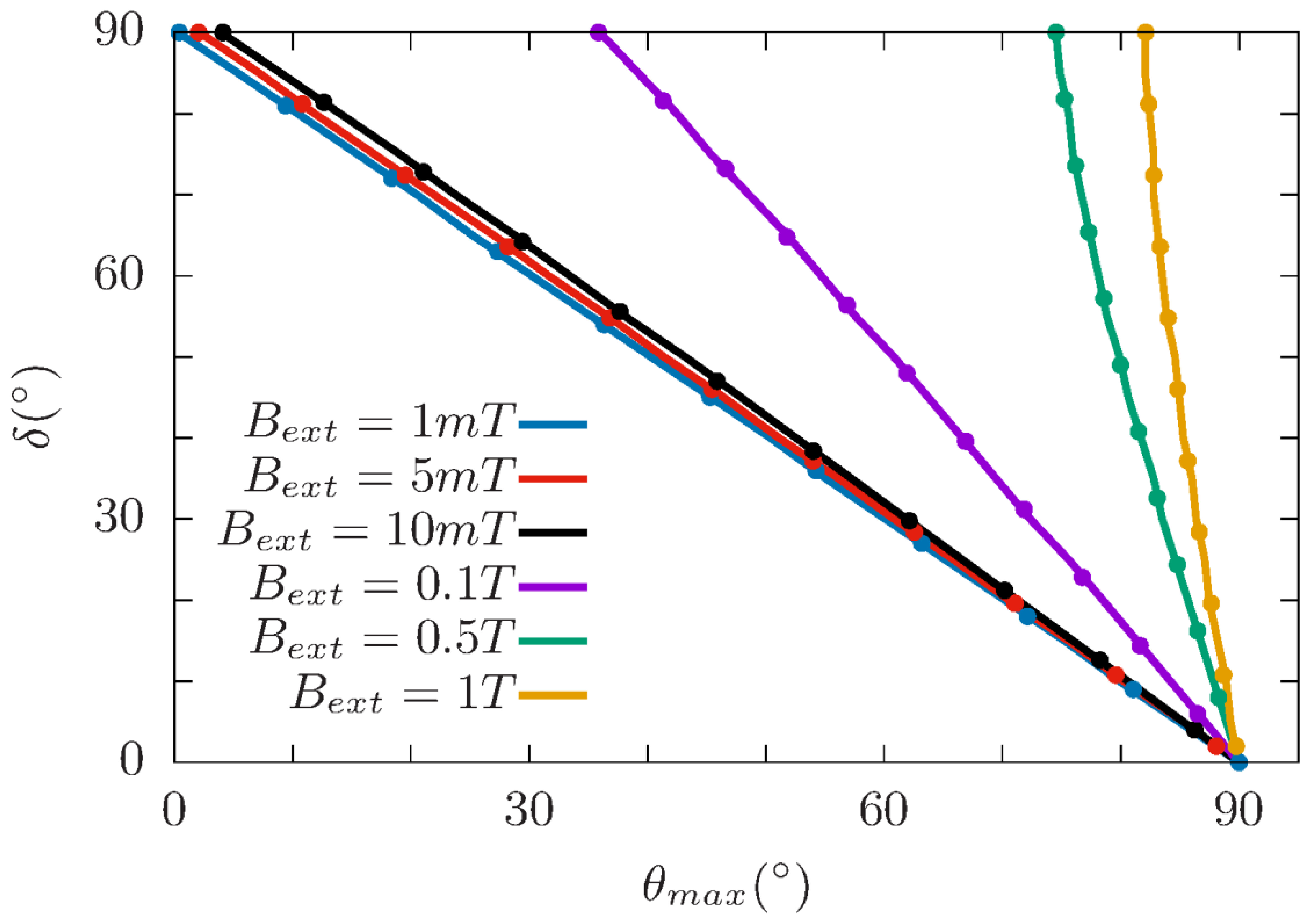}
        \caption{\label{figmax}
Detuning angle $\delta$ (see Eq. (\ref{def:delta})) as a function of $\theta_{max}$ (see Fig. 4) obtained with two methods: solid
lines show the results obtained from the exact diagonalization of the
few electron model while dots show results obtained from the
transcendental Eq. (\ref{thmax}). 
}
\end{figure}

\noindent In Fig. \ref{figmax}  we plot the anisotropy angle $\delta$ as a function
of $\theta_{max}$ for different values of the external magnetic
field. Solid lines and filled circles show an excellent agreement of the 
results obtained by numerical diagonalization of our multi-orbital model and the ones 
obtained  from solving Eq. (\ref{thmax}).
It is clear from this figure that in the small magnetic field limit 
($B_{ext}\rightarrow 0$) all the curves converge to the expression in Eq. (\ref{tad0}). This suggests that carrying out the experiments with small 
magnetic fields (it is even possible to carry them out at a vanishing field 
\cite{willke2019b}) would make it possible to estimate the misalignment angle $\delta$ 
without having to infer the values of $\varepsilon_{dip}$ and $\varepsilon_{exc}$.
We also note that, in this limit, the relation between $\delta$ and $\theta_{max}$ is independent of $V_{DC}$,
which should simplify the experiment.




\section{Summary and conclusions}
\label{sc}
In this work we have studied the  interactions between a  STM tip and a single Fe atom on an MgO surface,
relevant for single spin ESR-STM experiments\cite{Baumann2015,willke2018}.
The tip influences the surface spin via three different mechanisms: dipolar coupling,  exchange interactions, and the electric coupling that induces piezoelectric displacements of the  Fe ion, modulating both its spin interactions with the tip as well as its $g$ factor and crystal field parameters. 
All these interactions produce  shifts of the Fe energy levels and a modification of the resonance frequency $f$.  Therefore, in order to use the surface spin as a sensor for magnetometry, it is important to assess the tip contributions.  

We have discussed first the existence of a sweet spot at which the dipolar and exchange spin couplings cancel each other, the so called NOTIN point.  In figure 2 we have calculated the NOTIN point for $V_{DC}=0$. We have then discussed the influence of $V_{DC}$ on the  resonance frequency $f$.   We have calculated a phase diagram for the stability of $f$ with respect to variations in $V_{DC}$ as a function of both tip-atom distance $d_{tip}$ and $V_{DC}$.  We stress that two different regimes may be of practical interest. First, the regions in this phase diagram where the frequency is very stable with respect to voltage fluctuations. We have discussed how in this region decoherence due to both mechanical and electrical noise would be mitigated.  Second, regions in which large variations of the resonant frequency can be obtained by changing $V_{DC}$.  Whereas this would increase decoherence it would permit an electrical control of the resonant frequency that may be convenient in some situations. 

All of our results depend on $\delta$, the misalignment angle between the tip magnetization and the external magnetic field. We propose the following experimental protocol to measure $\delta$, based on measuring   $f(\theta)$, where $\theta$ is  the orientation of the applied field. We find that the maxima of this curve depend on $\delta$ and in the limit of vanishing external field, this relation is straightforward: these angles, $\theta$ and $\delta$, are dephased by $90^{\circ}$ (see Eq. (\ref{tad0})). 

These results may help the design of  future ESR-STM experiments, especially in cases where accurate
magnetometry is required \cite{yelko2022,Josehqd2022}. In this work, we have focused on the case of Fe adatoms on MgO.   Similar  results should be obtained in the  case of other ESR-STM active adatoms, such as Ti-H on MgO, with quantitative modification arising from the differences of  $g$ factor anisotropies and magnetic moments between these systems.

\section*{ACKNOWLEDGMENTS}

A.F. acknowledges ANPCyT (PICT2019-0654). A.F., S.S.G. and S.A.R.
acknowledge CONICET (PUE22920170100089CO and PIP11220200100170) and 
partial financial support from UNNE.
J.F.R.  acknowledges financial support from 
 FCT (Grant No. PTDC/FIS-MAC/2045/2021),
FEDER /Junta de Andaluc\'ia, 
(Grant No. P18-FR-4834), 
Generalitat Valenciana funding Prometeo2021/017
and MFA/2022/045,
and
 funding from
MICIIN-Spain (Grant No. PID2019-109539GB-C41).



\appendix

\section{Multi-orbital Electronic Model}\label{a1}

We start with the Hamiltonian for the 6 electrons in the $d$-levels of Fe:

\begin{equation}
{\cal H}_{\rm Fe}=H_{\rm CF}+H_{\rm SO}+ H_{\rm Z} + H_{ee}+H_{\rm Tip}
\label{aHCI}
\end{equation}
where $H_{\rm CF}$ is the crystal field Hamiltonian, $H_{\rm SO}$ is the spin-orbit coupling, $H_{\rm Z}$ is the Zeeman Hamiltonian,  $H_{ee}$ is
the Coulomb term and  $H_{\rm Tip}$ accounts for the interaction of the atom with the tip.
\noindent The Crystal Field part of the Hamiltonian is obtained from
the representation of the DFT Hamiltonian in the basis of maximally localized 
Wannier orbitals \cite{RevModPhys.84.1419,ferron2015,ferron2015b,lado2017}
\begin{equation}
H_{CF}=\frac{1}{2}\sum_{m,m',\sigma} \langle m|h_{\rm CF}|m'\rangle d_{m\sigma}^\dag d_{m'\sigma},
\label{hcf}
\end{equation}

\noindent where $d_{m\sigma}^\dag$ ($d_{m\sigma}$) denotes the creation
(annihilation) operator of an electron with spin $\sigma$  in the $\ell=2, 
\ell_z=m$ state of the Fe atom, denoted by  $\phi_{m}(\vec{r})$,
assumed to be equal to the product of a radial hydrogenic function (with  
effective charge $Z$ and a  effective Bohr radius $a_\mu$) and a
spherical harmonic. The one-particle elements are calculated from
\cite{lado2017}

\begin{equation}
h_{\rm CF}=D l_z^2+F \left(l_x^4+l_y^4\right)
\end{equation}

\noindent where $D=-290$ meV, and $F=-10$ meV are Crystal Field parameters 
obtained from DFT calculations and Wannierization \cite{lado2017,ferron2019}. We add the spin-orbit
coupling operator\cite{ferron2015,ferron2015b,lado2017}

\begin{equation}
	H_{\rm SO}=\lambda_{SO}\,\sum_{mm',\sigma\sigma'} \langle m\sigma|
	\vec \ell\cdot \vec S|m'\sigma'\rangle d_{m\sigma}^\dag d_{m'\sigma'},
	\label{hso}
\end{equation}

\noindent where we take $\lambda_{SO}=35$ meV \cite{lado2017,ferron2015}. Zeeman 
Hamiltonian reads:

\begin{equation}
\label{zeem}
H_{\rm Z}=\mu_B \vec B\cdot \sum_{mm',\sigma\sigma'} \langle m,\sigma |\left(\vec l+g\vec S\right)
|m'\sigma'\rangle d_{m\sigma}^\dag d_{m'\sigma'},
\end{equation}

\noindent where $g=2$. The Coulomb term reads:

\begin{equation}
H_{{\rm ee}}=\frac{1}{2}\sum_{m,m'\atop n,n'}
V_{mnm'n'}
\sum_{\sigma\sigma'}d_{m\sigma}^\dag d_{n\sigma'}^\dag d_{n'\sigma'}d_{m'\sigma},
\label{hcoul}
\end{equation}

\noindent For the evaluation of the Coulomb integrals $V_{mnm'n'}$  we
transform the angular part to a basis of eigenstates of $l=2$.
In the basis of eigenstates of
$l=2$, all the Coulomb integrals scale linearly with the value
of $V_{0000}=U$\cite{ferron2015,ferron2015b}. Here we take, as in
previous works, $U=5.0$ eV \cite{lado2017,ferron2015}. Now, to end with an adequate description of the situation, we need to describe the interaction of the atom with the tip. Along this work, we  consider exchange and dipolar interaction $H_{Tip}=H_{\rm ex}+H_{\rm dip}$. We can write the exchange interaction as $J_{ex}(z)\vec{S}_{tip}\cdot \vec{S}$ \cite{lado2017} where $S_{tip}$ is the total spin of the spin-polarized tip and $J_{ex}$ is the exchange coupling between the surface spin and the tip. The exchange coupling depends exponentially on the tip-adatom distance and it can be written \cite{lado2017}:

\begin{equation}
 J_{ex}(z)=J_{0}\,e^{-z/l_{0}}\,.
\end{equation}

\noindent where $J_0$ and $l_0$ depend on the tip and the adatom. The exchange 
coupling can be written as in most of the experimental works\cite{willke2019} 
as $J_{ex}(z)=J^{\ast}_{0}\,e^{-(z-z_{pc})/l_{0}}$ by doing 
$J_{0}=J^{\ast}_{0}e^{z_{pc}/l_{0}}$ and $z_{pc}$ is the tip heights above 
point contact and can be measured for Ti ($z_{pc}\simeq 0.3$ nm) and Fe 
($z_{pc}\simeq 0.4$ nm) atoms \cite{willke2019}.  We ignore the quantum 
fluctuations of the magnetic moment of the apex atom or atoms, quenched by the 
combination of an applied magnetic field and strong Korringa damping with the 
tip electron bath\cite{Delgado_Hirjibehedin_sc_2014}. Therefore, we treat the tip spin in a mean-field or 
classical approximation, following [\onlinecite{Yan_Choi_natnano_2015,
lado2017}], and replace $S_{tip}$ by its statistical average  
$\langle S_{tip}\rangle$. Then, the tip-atom exchange interaction contribution 
to the Hamiltonian reads
\begin{equation}
\label{ex1}
H_{\rm ex}=J_{ex}\langle\vec{S}_{tip}\rangle\cdot \sum_{mm',\sigma\sigma'} \langle m,
\sigma |\vec S
|m'\sigma'\rangle d_{m\sigma}^\dag d_{m'\sigma'},
\end{equation}

\noindent where $\braket{\vec{S_{tip}}}=\braket{S_{tip}}\hat{n}_T$
and $\hat{n}_T=(\cos(\theta+\delta),0,\sin(\theta+\delta))$ (see Fig. \ref{fig1}). 

The dipolar interaction between the magnetic moment of the tip and the surface spin, where the tip creates a magnetic field whose orientation depends on the tip characteristics, gives us a dipolar term of the form:

\begin{equation}
\label{fip1}
H_{\rm dip}=\mu_b\vec{B}_{dip}\cdot \sum_{mm',\sigma\sigma'} \langle m,
        \sigma |\left(\vec l+g\vec S\right)
|m'\sigma'\rangle d_{m\sigma}^\dag d_{m'\sigma'},
\end{equation}

\noindent where the effective magnetic field created by the tip is defined in Eq. (\ref{eq:dip}).

The few-body Hamiltonian \ref{aHCI}
is solved exactly, by numerical 
diagonalization in a space constructed with all the states that accommodate 
six electrons in five spin-degenerate $d$ orbitals. 

\section{Effective spin model}\label{a2}

The lowest energy manifold of the Hamiltonian defined in the previous section has five states separated from the rest of the spectra. These states correspond to a ground state with $S=2$ result of accommodating 6 electrons in 5 $d$ orbitals and can be described in terms of a simple effective spin model \cite{ferron2015,ferron2019}:

\begin{eqnarray}\label{hspin}
	\hat{H}_{eff}=-{\cal D}_2\hat{S}_z^2+{\cal D}_4\hat{S}_z^4-
	{\cal F}(\hat{S}_+^4+\hat{S}_-^4)+{\cal H}_1\,.
\end{eqnarray}

\noindent where the spin operators act on the $S=2$ subspace. The term ${\cal H}_1$
accounts for the interaction of the adatom with the tip and the external 
field (see Eq. (\ref{bej}) and Eq. (\ref{h1j})). The anisotropy
terms ${\cal D}_2$, ${\cal D}_4$ and ${\cal F}$ can be obtained from the
diagonalization of Hamiltonian Eq. (\ref{aHCI}) with no magnetic field and no tip. We obtain ${\cal D}_2=4.86$ meV, ${\cal D}_4= 0.23$ meV, and
${\cal F}=4.06$ neV. Now, the spectrum of the ground state manifold has an EPR active space formed by a doublet of states with $S_z=\pm 2$. Yet, this doublet has a zero-field splitting (ZFS), given by 
$\Delta= 48{\cal F}=0.2$ $\mu$eV, due to quantum spin tunneling 
\cite{Klein_ajp_1951,garg1993topologically,Delgado_Loth_epl_2015}. 

Our model for the Fe atom at MgO taking into
account the interaction of the surface spin with the magnetic moment of the
tip, described by Hamiltonian (\ref{aHCI}), can be solved by numerical
diagonalization. We can calculate the resonance frequency $f=\Delta/h=
(E_1-E_0)/h$ as a function of the tip-atom distance $z=d_{tip}$ for different values of $J_0$ and $M_{tip}$. Comparing with experimentally determined parameters and the resonance frequency curves obtained
experimentally  \cite{willke2018,willke2018b,willke2019b,willke2019,seifert2019,
seifert2021,seifert2020}, in particular paying attention to the results
presented in [\onlinecite{willke2019}], [\onlinecite{seifert2020}] and
[\onlinecite{seifert2021}], a reasonable set of values are 
$l_0=0.04-0.06$ nm,  $J_0=10-60$ eV
($J^\ast_0=0.5-3$ meV) and $M_{tip}=10\mu_B-40\mu_B$.

\subsection*{Perturbative expresions}

The eigenvalues of the effective spin Hamiltonian assuming  $B_{eff}^{(x)}=0$  can be written

\begin{eqnarray}
 &&E_{4}=0\,,  \nonumber \\ 
	&&E_{3}=-{\cal D}_2+{\cal D}_4+B_{eff}^{z}\,,  \nonumber \\ 
	&&E_{2}=-{\cal D}_2+{\cal D}_4-B_{eff}^{z}\,,  \nonumber \\ 
	&&E_{1}=-4{\cal D}_2+16{\cal D}_4+\sqrt{(24{\cal F})^2+
	(2B_{eff}^{z})^2}\,,  \nonumber \\ 
	&&E_{0}=-4{\cal D}_2+16{\cal D}_4-\sqrt{(24{\cal F})^2+
	(2B_{eff}^{z})^2}\,,  \nonumber 
\end{eqnarray}

\noindent and the eigenvectors

\begin{eqnarray}
 &&\ket{4}=\ket{S_z=0}\,, \nonumber \\
 &&\ket{3}=\ket{S_z=1}\,, \nonumber \\
 &&\ket{2}=\ket{S_z=-1}\,, \nonumber \\
 &&\ket{1}=C_1^{1}\ket{S_z=2}+C_2^{1}\ket{S_z=-2}\,, \nonumber \\
 &&\ket{0}=C_1^{0}\ket{S_z=2}+C_2^{0}\ket{S_z=-2}\,, \nonumber 
\end{eqnarray}

\noindent where $\{\ket{S_z=i}\}$ are eigenstates of $S_z$ 
for $S=2$. We introduce $B_{eff}^{x}$ as a perturbation. Using perturbation up to second order in $B_{eff}^{x}$, we have that

\begin{equation}
 E_{i}^{(2)}=E_{i}^{(0)}+{\left(\mu_B g^x 
 B_{eff}^{x}\right)}^2\,\sum_{m\neq i}^{4}\,\frac{|\braket{m|\hat{S}_x|i}|^2}{E_{i}^{(0)}-E_{m}^{(0)}}\,.
\end{equation}

\noindent Assuming  $B_{eff}<<{\cal D}_i$ with $i=2,4$ we obtain

\begin{eqnarray}\label{ecuation_2}
	&& E_{i}^{(2)}=E_{i}^{(0)}+\frac{3}{2}\,\frac{{\left(\mu_B g^x 
 B_{eff}^{x}\right)}^2\,}{E_{i}^{(0)}-E_{3}^{(0)}}\,.
\end{eqnarray}
We thus see that, in the subspace of the two lowest energy states of the Fe adatom, the effect
of $B_x$ is negligible, to linear order in $B_x$.
\noindent and finally we can write the gap of the system as follows
\begin{equation}\label{gapap}
 \Delta \approx 2\sqrt{(24{\cal F)}^2+(2\mu_B g^z B_{eff}^{z})^2)}
\end{equation}

As the anisotropy terms, we obtain the $g$ tensor 
from the diagonalization of Hamiltonian Eq. (\ref{aHCI}). For the values quoted above and ignoring the effect of the tip, we get $g_z=2.8$.

\section{Effect of the external electric field on the Hamiltonian parameters}\label{a4}


Our model assume a piezoelectric distortion of the atom-tip distance due to the external electric field. This distortion 
modulates crystal field parameters in the Hamiltonian Eq. ($\ref{aHCI}$):

\begin{eqnarray}\label{dyf}
 && F=F_{eq}+\frac{dF}{dz}\bigg|_{z_{eq}}\delta z \,, \nonumber \\
 && D=D_{eq}+\frac{dD}{dz}\bigg|_{z_{eq}}\delta z \,,
\end{eqnarray}

\noindent and tip terms directly by their dependency with $\delta z$ through $z=d_{tip}-\delta z$ where $d_{tip}$ is the tip-atom distance at the equilibrium. Here we take, using DFT calculations\cite{lado2017},
$\frac{dF}{dz}\bigg|_{z_{eq}}=280\frac{meV}{nm}$ and
$\frac{dD}{dz}\bigg|_{z_{eq}}=0$. 

For the effective spin Hamiltonian Eq. (\ref{hspin}) and Eq. (\ref{hspin-g}) we can calculate
the modulation of the parameters by direct diagonalization of Eq. (\ref{aHCI})
taking into account the modulation of the crystal field parameters described
above:

\begin{eqnarray}\label{spin_parameters_mod}
 && {\cal D}_2={\cal D}_{2_{eq}}+\alpha_{{\cal D}_2}\,\delta z \nonumber\\
 && {\cal D}_4={\cal D}_{4_{eq}}+\alpha_{{\cal D}_4}\,\delta z \nonumber\\
 && {\cal F}={\cal F}_{eq}+\alpha_{{\cal F}}\,\delta z \nonumber\\
 && g^x=g^{x}_{eq}+\alpha_{g^x}\,\delta z \nonumber\\
 && g^z=g^{z}_{eq}+\alpha_{g^z}\,\delta z\,.
\end{eqnarray}

\noindent with

\begin{equation}
\alpha_X=\frac{\partial X}{\partial F}\,\frac{\partial F}
{\partial z}\bigg|_{z_{eq}}+\frac{\partial X}{\partial D}\,
\frac{\partial D}{\partial z}\bigg|_{z_{eq}}.
\end{equation}

Finally, calculated modulation and equilibrium parameters for the effective Hamiltonian are depicted in Table \ref{t1}.



\begin{table}
  \begin{center}
\begin{tabular}{|c|| c | c|} 
 \hline
    $X$       & $X_{eq}$      & $\alpha_X$ \\
    \hline 
     \hline
  ${\cal D}_2$&     $4.9$ meV & $134$ $\mu$eV/pm\\
  ${\cal D}_4$&     $0.2$ meV & $14$ $\mu$eV/pm \\
  ${\cal F}$&     $4.1$ neV & $-0.0002$ $\mu$eV/pm \\
  $g^x$        &     $2.0$     & $-0.36$ nm$^{-1}$ \\
  $g^z$        &     $2.8$     & $-9.02$ nm$^{-1}$ \\
  \hline
\end{tabular}
\caption{\label{t1} Parameters of the effective Hamiltonian and their modulations.}
\end{center}
\end{table}

\bibliography{biblio}{}

\end{document}